\documentclass[sigconf]{acmart}
\usepackage{graphicx}

\AtBeginDocument{%
  }

\setcopyright{acmcopyright}
\usepackage{hyperref}
\usepackage{fancyhdr}

\copyrightyear{2018}
\acmYear{2018}
\acmDOI{XXXXXXX.XXXXXXX}

\begin{document}

\title{Autonomous Agents in Software Development: A Vision Paper}

\author{Zeeshan Rasheed, Muhammad Waseem, Kai-Kristian Kemell, Wang Xiaofeng, Anh Nguyen Duc, Kari Systä, Pekka Abrahamsson}
\affiliation{%
  \institution{Faculty of Information Technology and Communication Science, Tampere University \\ Faculty of Information Technology, Jyväskylä University, Jyväskylä, Finland \\ Faculty of Mathematics and Natural Sciences, University of Helsinki, Finland \\ Faculty of Engineering, Free University of Bozen Bolzano, Italy}
  \country{Finland and Italy}
}
\email{zeeshan.rasheed@tuni.fi}
\email{muhammad.m.waseem@jyu.fi}
  











\pagestyle{fancy}
\fancyhf{} 
\fancyfoot[C]{\thepage} 
\renewcommand{\headrulewidth}{0pt} 

\begin{abstract}
Large Language Models (LLM) and Generative Pre-trained Transformers (GPT), are reshaping the field of Software Engineering (SE). They enable innovative methods for executing many software engineering tasks, including automated code generation, debugging, maintenance, etc. However, only a limited number of existing works have thoroughly explored the potential of GPT agents in SE. This vision paper inquires about the role of GPT-based agents in SE. Our vision is to leverage the capabilities of multiple GPT agents to contribute to SE tasks and to propose an initial road map for the future work. We argue that multiple GPT agents can perform creative and demanding tasks far beyond coding and debugging. GPT agents can also do project planning, requirements engineering, and software design. These can be done through high-level descriptions given by the human developer. We have shown in our initial experimental analysis for simple software (e.g., Snake Game, Tic-Tac-Toe, Notepad) that multiple GPT agents can produce high-quality code and document it carefully; We argue that it shows a promise of unforeseen efficiency and will dramatically reduce lead-times. To this end, we intend to expand our efforts to understand how we can scale these autonomous capabilities further. We provide the video link below as a demonstration of our initial work \footnote{\url{https://www.youtube.com/watch?v=GbOJP5rrSjI}}.

\end{abstract}

\keywords{OpenAI, AutoGPT, Artificial Intelligence, Natural Language Processing, Generative AI, Software Engineering}

\maketitle

\section{Introduction}
\label{Introduction}
AI has reshaped our interaction with machines and impacted many industries. Its most promising use is in Natural Language Processing (NLP), which helps computers understand and interact with human language \cite{chowdhary2020natural}. Recent advancements in NLP have led to the development of powerful language models such as the GPT series \cite{radford2018improving, radford2019language, ouyang2022training, brown2020language}. These GPT models have significantly advanced the field of NLP by demonstrating remarkable capabilities in generating coherent and contextually relevant text \cite{topal2021exploring}. Pre-trained on extensive volumes of text data, these models have exhibited exceptional performance across a wide range of NLP tasks, including language translation, text summarization, and question-answering. Specifically, the ChatGPT model has demonstrated its potential in various domains such as education, healthcare, logical reasoning, text generation, code generation, human-machine interaction, and scientific research.

The interaction between GPT models and the domain of Software Engineering (SE) has led to significant advancements and intriguing possibilities \cite{treude2023navigating}. GPT, renowned for its expertise in NLP, has found applications in various aspects of SE. These range from assisting in code generation and auto-completion to aiding in documentation, bug detection, and software requirement analysis. GPT models have shown their potential to streamline multiple facets of the software development lifecycle \cite{khan2022automatic}. The integration of GPT's language comprehension capabilities with SE tasks holds the promise of enhancing developer productivity, enabling rapid prototyping, and fostering innovative problem-solving approaches \cite{ma2023scope}. The software development landscape is experiencing a significant transformation due to the integration of autonomous GPT agents. These advanced language models have emerged as a formidable force, leveraging the power of AI to autonomously generate code, offer insights, and assist developers across various stages of the software development lifecycle. However, challenges such as ensuring code correctness, maintenance, and bridging the gap between natural language and programming semantics remain critical considerations in this symbiotic relationship.

This paper presents the interactions between GPT models and SE, and investigates their applications, benefits, and complexities in autonomously generating various SE tasks. Our vision leans more toward software teams in which multiple GPT agents play an important role in automating various SE tasks. In this vision paper, we demonstrate the success of our initial work in autonomous code generation, which shows remarkable efficiency and reduced development time. Our initial results suggest that the thoughtful implementation of GPT agents can revolutionize the development lifecycle but necessitates a re-evaluation of team roles and collaboration strategies for effective implementation. With this goal in mind, we plan to extend this work to encompass autonomous code debugging, maintenance, and documentation. By capitalizing on their natural language understanding and generation capabilities, these agents will excel at error identification, contextual understanding, automated reporting, and code fixes, all while continuously learning and adapting to the specific needs of the software project. By introducing a diverse array of specialized agents and fostering collaborative interactions, this framework will empower software development teams to expedite project timelines, improve code quality, and enhance software reliability. This vision paper lays the foundation for further research and development, offering a glimpse into a future where coding becomes a more intuitive, efficient, and inclusive process. 

\section{Background}
\label{Background}

\subsection{Generative AI}
\label{Generative AI}
Generative AI refers to a category of AI models and algorithms that are designed to generate new content that is often similar to content created by humans \cite{baidoo2023education}. 
This type of AI has experienced notable progress in recent times \cite{cao2023comprehensive}. Nowadays, generative AI has been utilized in various fields, such as NLP, computer vision, and image and video generation \cite{hacker2023regulating}. In NLP, generative AI techniques are commonly used for various tasks, including text generation, machine translation, dialog systems, and code generation. According to Aydn \textit{et al}. \cite{aydin2023chatgpt}, researchers have leveraged generative AI models to improve various NLP tasks. Generative Adversarial Networks (GANs) and autoregressive language models, such as GPT, are subsets of generative AI that have been applied to tasks like text generation, machine translation, and dialogue systems.

The GPT model is a specific type of generative AI model that excels at generating human-like text due to its architecture, pre-training, and fine-tuning processes \cite{radford2018improving}. It shows the capabilities of generative models in the domain of NLP, and its success has spurred further research and development in text generation and language understanding \cite{liu2023gpt}. The foundation of the GPT model can be traced back to the introduction of the transformer architecture proposed by Vaswani \textit{et al}. \cite{vaswani2017attention}. This architectural innovation transformed the field of NLP by introducing the self-attention mechanism, enabling the model to capture contextual connections among words, irrespective of their position within a sequence. In 2018, OpenAI introduced the GPT-1 model to demonstrate the potential of large-scale language models for text generation tasks \cite{radford2018improving}. 
Progress in GPT models has increased efficiency, adaptability to various tasks, and the potential for practical applications in various industries. Several researchers and OpenAI have made significant contributions to improving the performance of GPT models by using a variety of techniques and approaches \cite{radford2018improving}, \cite{radford2019language}, \cite{ouyang2022training}, \cite{brown2020language}. For instance, adding more parameters to the GPT model allows it to capture more complex patterns and nuances in language. Models like GPT-2, GPT-3, and GPT-4 are notable examples of scaling, with billions of parameters \cite{peng2023instruction}, \cite{dale2021gpt}, \cite{budzianowski2019hello}. According to White \textit{et al}. \cite{white2023chatgpt}, GPT models, known for their NLP capabilities, show significant potential in transforming different aspects of the software development process. 

\subsection{GPT Models in SE}
\label{GPT models in SE}
Our research's main goal is to help the SE research community benefit from the important motivators and ethical principles of using the GPT model in SE research. GPT models have shown promise in various SE applications \cite{feng2023investigating}. They have been trained on large code repositories, which enables them to generate code snippets or even entire programs based on natural language prompts \cite{floridi2020gpt}. As Treude \textit{et al}. \cite{treude2023navigating} mentioned, the GPT model and SE are related by applying NLP techniques to various tasks within the software development lifecycle. GPT's language generation capabilities offer valuable assistance and enhancements to SE processes \cite{thiergart2021understanding}, \cite{hornemalm2023chatgpt}. By utilizing the remarkable natural language generation capabilities of GPT models, various SE tasks can now be automated and streamlined, including code generation, error detection, documentation creation, and beyond. Through GPT-powered code completion and generation, developers can swiftly produce high-quality code and even entire programs, significantly expediting the software development lifecycle \cite{dong2023self}. Ma \textit{et al}. \cite{ma2023scope} and Nascimento \textit{et al}. \cite{nascimento2023comparing} conducted comprehensive empirical studies to investigate the capabilities of GPT model for different SE tasks. 


Several studies (e.g., \cite{ahmad2023towards}, \cite{barke2023grounded},\cite{vaithilingam2022expectation}) has explored GPT models in SE. However, much of this research has been limited to case studies, with a focus on user perceptions in coding and writing. To fill this gap, our goal is to leverage multiple GPT AI agents for autonomous tasks such as code generation, maintenance, debugging, and documentation based on developer-provided high-level descriptions. This approach aims to streamline software development, reduce development time, and tackle key challenges in conventional processes.

\section{A Collaboration Framework of Multiple GPT-agents}
\label{conceptual framework}

Integrating multiple GPT agents into a unified framework enhances AI's problem-solving prowess. Our goal is a multi-GPT agent SE framework that streamlines software development, maintenance, bug detection, and documentation. The framework autonomously handles these tasks, improving efficiency and reducing development time. Our initial tool, employing multiple AI agents for code generation, demonstrates significant time savings.

Researchers have increasingly embraced AI systems for automating SE tasks \cite{zong2022survey}, particularly in code generation and test strategy formulation, as highlighted by Nascimento \textit{et al}. \cite{nascimento2023comparing}. However, shortcomings persist in testing and debugging, as emphasized by Liu \textit{et al}. \cite{liu2023your}. Our core objective is to introduce an encompassing framework where multiple GPT agents collaborate to reshape the software development lifecycle. Leveraging their collective capabilities, we aim to automate high-level code generation, debugging, maintenance, testing, and documentation, effectively reducing development time and enhancing overall efficiency. In the following sections, we discuss key research challenges that have emerged in our quest to explore autonomous software development, debugging, and maintenance and their impact on the software industry and SE research.

\paragraph{\textbf{Autonomous code generation}}:
GPT-based agents offer a range of capabilities within SE due to their natural language understanding and generation capabilities \cite{radford2018improving}. Recently, automated code generation has made significant advancements in both academic and industrial domains \cite{shen2022incorporating}, \cite{dong2023codep}. However, generating code for complex tasks is still challenging \cite{dong2023self}. To this end, we utilize multiple GPT-based agents to automatically generate code or entire software components based on high-level descriptions. This can accelerate the software development process by automating routine coding tasks. Our Future goal is to investigate methods for automatically designing and optimizing prompts for code generation tasks, design a better evaluation metric to assess the correctness of the generated code automatically, and further assess and improve the code generation quality. We also plan to explore more and better test generation techniques to keep improving the benchmark quality.

\paragraph{\textbf{Autonomous code debugging}}:
Leveraging multiple GPT agents to establish autonomous code debugging capabilities within the domain of SE can be a transformative approach to streamline and enhance the software development process. Recently, there has been a growing trend in using large language models to detect bugs in code, which opens the door to several new scenarios and prospects in software development \cite{vaithilingam2022expectation}. ChatGPT has demonstrated unprecedented capabilities in understanding, generating, and debugging code \cite{tian2023chatgpt}, although it cannot replace traditional debugging practices. According to Buscemi \textit{et al}. \cite{buscemi2023comparative}, there is a lack of a comprehensive framework for autonomous code debugging, which can help developers automate error reports and fix bugs. To this end, we propose a framework that will help developers autonomously identify and fix the bug in code. We will employ multiple GPT agents to monitor the codebase continuously. These GPT agents will possess the natural language understanding and pattern recognition capabilities to pinpoint errors and issues in the code. 

\paragraph{\textbf{Autonomous code maintenance}}:
Automatic code maintenance can help evaluate code repositories to provide meaningful insight into code quality and also help improve code comprehension \cite{jiang2021cure}. In this project, we will utilize multiple GPT agents to build a framework for autonomous code maintenance. Multiple GPT agents will continuously monitor software systems. Each agent will focus on specific aspects of the codebase, such as security vulnerabilities, performance bottlenecks, or bug detection. Once an issue is identified, the GPT agents will generate context-aware solutions. Our proposed model will understand the problem and propose fixes or optimizations considering the software's architecture, coding standards, and best practices. The proposed model will automatically generate code fixes or optimizations. This includes modifying the code to resolve bugs, enhance security, or improve performance. By automating the code-fixing process, these agents reduce the need for manual intervention, speeding up maintenance tasks and reducing the risk of introducing new errors.

\section{Preliminary Results}
\label{preliminary result}

In this section, we present preliminary results from our work in developing a novel tool that utilizes the capabilities of eight GPT-based agents to autonomously generate code from high-level descriptions provided by developers. These agents, each specializing in a distinct aspect of the software development process, collectively contribute to improving efficiency and enhancing the software development life cycle. Below, we discuss how eight GPT agents collaboratively work to autonomously generate code based on the given description.

\begin{figure}
    \centering
    \includegraphics[width=0.4\textwidth]{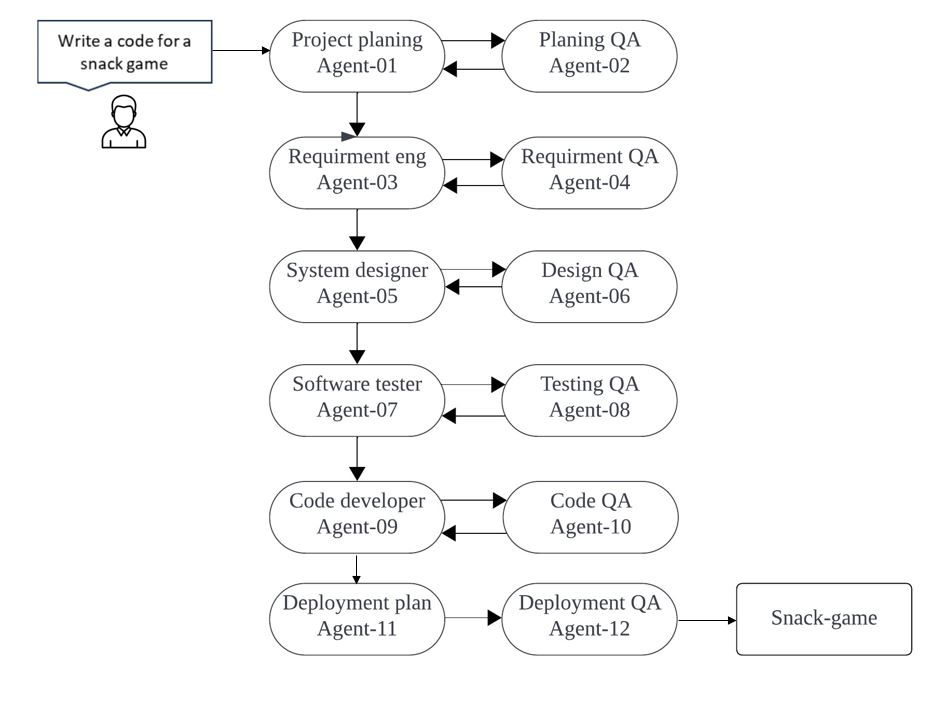}
    \caption{Autonomous code generation for snake-game}
    \label{GPT agents}
\end{figure}

Figure \ref{GPT agents} demonstrates the collaborative efforts of several GPT-based agents working together to generate complex code autonomously. 
Our Agent-1 specializes in project planning, which involves defining the software project's scope, goals, and objectives and creating a detailed plan for how the project will be executed. The primary role of Agent-1 is to define and lay out the entire project's scope, objectives, and the steps required to achieve those objectives. Agent-2, responsible for project planning quality analysis, focuses on ensuring the quality and effectiveness of the project plan created by Agent-1. Agent-2 assesses whether the project plan is comprehensive, feasible, and aligns with the organization's goals.
The next step is requirement engineering, Agent 3 demonstrated the ability to extract and interpret high-level requirements effectively. This fundamental stage guarantees that the following development phases are rooted in clearly understanding the project's objectives and constraints. Collaborating with Agent-3, the requirement quality analysis Agent-4 plays an important role in assessing the clarity of the extracted requirements. Early results suggest a significant improvement in the quality and consistency of requirements documentation.
The system design Agent-5 contributes to the architectural planning of the software, effectively transforming high-level requirements into structured system blueprints. Our initial findings indicate that this agent accelerates the design phase while aligning with project specifications.
Agent-6 conducts design quality analyses parallel to system design, identifying potential issues and inconsistencies in the developed system architecture. The timely identification of design flaws has already led to significant savings in time and resources. Agent-5 expert in generating test scripts based on the developed code. This automated testing capability significantly reduces the manual effort required for quality assurance.
Collaborating closely with Agent-7, the testing quality analysis Agent-8 assesses the effectiveness of the automated test suites generated. Initial evaluations suggest an increased efficiency in identifying and fixing software defects. Our software development agent-9, contributes to generating code based on high-level design specifications. Early performance metrics indicate a notable reduction in development time while maintaining code quality standards. Agent-10, responsible for code quality analysis, assesses the generated code to ensure it complies with coding standards and best practices. Preliminary results demonstrate its efficacy in ensuring the production of clean and maintainable code.
Finally, Agent-11 plays a crucial role in planning the deployment phase of a software project. Deployment planning involves the coordination and execution of activities necessary to release the software to the production environment or to end-users. Ultimately, Agent-12 reviews, assesses, and improves the quality of the deployment plan created by Agent-11 to ensure a smooth and reliable deployment process.

The agents were prompted to "Develop a snakegame". As the outcome of the autonomous development process, the agents (run by GPT3.5 Turbo) produced a requirements engineering specification, a software design plan, commented software code, a test and deployment plans. All material produced were reviewed and revised. The amount of documentation including review rounds is 6216 words. This is about 13 pages of single-spaced text. Requirements specification agent produced 7 functional requirements, 7 non-functional requirements and 4 constraints. 11 were fully met, 2 partially, 4 not verified and 4 were not met. There was no restart-option, the code was not  commented according to the standard PEP 257 and object-oriented programming principles were not implemented. The code did not include test-cases. The actual generated software code was 95 lines of code and the whole project took less than 4 minutes to complete. Game required human debugging to make it run.

The second experiment was run by GPT-4. The amount of documentation including review rounds is 4565 words. Requirements specification agent produced 9 functional requirements and 9 non-functional requirements and 0 constraints. 14 were fully met and 4 were not met. Game area was not fixed as requested, score display was not all the time visible, missing pause/resume function and missing user interface. The code does not include test-code. The actual generated software code was 75 lines and was a bit slower to produce than the first experiment. Game was able to run directly after code generation.

When the two sets of requirements are compared, it looks like the GPT-3.5 outperforms GPT-4 by a small margin. Against our expectations, GPT-3.5 generated requirements that were higher quality because they were  more complete, specific, and provided a more comprehensive foundation for both development and testing. As an example, GPT-4 did not produce a game with graphical user interface.

The third test was run by GPT-4 and the prompt was changed to "develop a snakegame with GUI". The complete documentation is 5366 words. There are 9 functional, 6 non-functional, 2 performance, 2 security requirements and 3 constraints. Only 6 requirements were fully met while 11 were not met. The remaining were undetermined. 94 lines of code was developed. Game worked out of the box but there was no possibility to be killed.

While these results are preliminary, they underscore the potential of our multi-agent framework in automating and optimizing various stages of the software development process. The most pressing issues are to understand how far we can scale up the autonomous capabilities, and where are the actual limits. Currently, the limits are connected to LLM memory. Further, our future goal is to extend this work to encompass autonomous code debugging, maintenance, and documentation by adding more GPT agents.
\section{Potential Impact}
\label{potential impact}
Using multiple GPT agents for the autonomous generation of code and entire software programs based on high-level descriptions provided by developers has the potential to revolutionize the software development landscape. This proposed model shows impressive efficiency gains in code generation, and we plan to extend this work to autonomous debugging processes and software maintenance. This will offer substantial benefits to the field of SE. The potential impact of this work is multi-faceted.

\textbf{Enhanced productivity}: Through the automation of SE tasks like code generation, testing, maintenance, and documentation, GPT-based agents liberate human developers to focus on their core strengths: creative problem-solving and innovation. Enhanced productivity can result in software development teams that operate more efficiently, enabling them to tackle larger and more intricate projects. Developers can swiftly translate high-level concepts into functional code, saving time and effort. This heightened productivity can lead to faster software delivery and reduced development costs.

\textbf{Improved code quality}: Automated testing, debugging, and maintenance performed by GPT-based agents can significantly influence software quality. The ability of these agents to identify and fix errors can significantly reduce software defects and enhance system reliability, leading to greater user satisfaction and trust.

\textbf{Scalability and adaptability}: The flexibility and scalability of GPT-based agents enable to accommodate projects of varying sizes and levels of complexity. This adaptability means thet both small startups and large enterprises can leverage these agents to enhance their software development processes.

\textbf{Streamlined debugging}: The model's ability to understand high-level descriptions aids in producing more coherent and logically structured code. Developers can identify and fix issues more efficiently, leading to improved code quality and quicker problem resolution.

\textbf{Reduced human error}: By automating code generation and debugging to a significant extent, the potential for human error is reduced. The model's ability to provide context-aware code suggestions and identify huge contributes to more accurate code development and problem-solving.

\textbf{Continuous learning and adaptation}: The continuous evolution and enhancement of GPT agents within this context actively participate in an ongoing learning process. As the agents encounter and fix a broader range of issues, their ability to assist in code maintenance and debugging tasks further improves.


\section{Conclusions}
\label{conclusions}
This paper proposes to explore how multiple GPT agents can perform various SE tasks, requirements analysis, design, code generation, debugging, and managing software maintenance autonomously. Through initial experiments, we have shown some interesting results which may have significant consequences. 

We demonstrate that our method significantly reduces development time and advances code generation methodologies, reinforcing the potential of AI-driven practices in the SE domain. With this goal in mind, we aim to extend our work to study how far we can scale these methods and where do really need human developers to be involved in. This will further push the limits of what is achievable in software engineering, making software development more efficient, accessible, and innovative.

Experimenting using our proposed framework, we can gain many insights on how AI technology can fundamentally change how software is developed. The various SE roles and activities are designed based on human nature, capabilities and limitations in knowledge, skills and communication. AI agents may not have these limits. Then a follow up question to ask is, do we still need the same activities and life cycles as we did in the past 50 to 60 years of software engineering?

\bibliographystyle{ACM-Reference-Format}
\bibliography{acmart} 

\end{document}